\documentclass[runningheads]{llncs}

\usepackage{eccv}

\usepackage{eccvabbrv}

\usepackage{graphicx}
\usepackage{booktabs}

\usepackage[accsupp]{axessibility}  

\usepackage{algorithm}
\usepackage{algpseudocode}

\usepackage{bbm}
\usepackage{nccmath}
\usepackage{amssymb} 
\usepackage{wrapfig}
\usepackage{multirow}
\usepackage{caption}
\usepackage{enumitem}

\usepackage{hyperref}

\usepackage{orcidlink}

\begin{document}

\title{OASIS: Online Adaptive Video Compression via Closed-loop Feedback Control} 


\author{Rentao Wan\orcidlink{0000-0003-4146-9684} \and
Jinho Park\orcidlink{0000-0003-1613-0769} \and
Mingoo Seok\orcidlink{0000-0002-9722-0979}}

\authorrunning{R.~Wan et al.}

\institute{Department of Electrical Engineering, Columbia University, NY 10027, USA 
\email{\{rw2899, jp4327, ms4415\}@columbia.edu}}

\maketitle

\begin{abstract}
  Computer vision systems are a key building block in an autonomous vehicle, responsible for a range of perception tasks. However, they incur massive data transmission over long communication links from multiple cameras, creating a critical bandwidth and energy bottleneck. Although conventional codecs such as H.264 can reduce data rates, they are ill-suited for real-time vision systems due to high processing latency and energy consumption, as well as their reliance on static user-defined compression settings.
  In light of these challenges, we propose OASIS, an adaptive video compression framework that integrates lightweight in-sensor compression with task-aware compression ratio control. Based on the real-time task performance, it dynamically updates the optimal compression ratio. Experimental results demonstrate that OASIS generalizes across multiple vision tasks, achieving on average a 6× data compression, 5.8× reduction in link power consumption, and 2.5× reduction in link latency, with at most 1.5\% performance degradation.

\end{abstract}

\section{Introduction}
\label{sec:intro}

\begin{figure}[tb]
  \centering
  \includegraphics[width=0.95\linewidth]{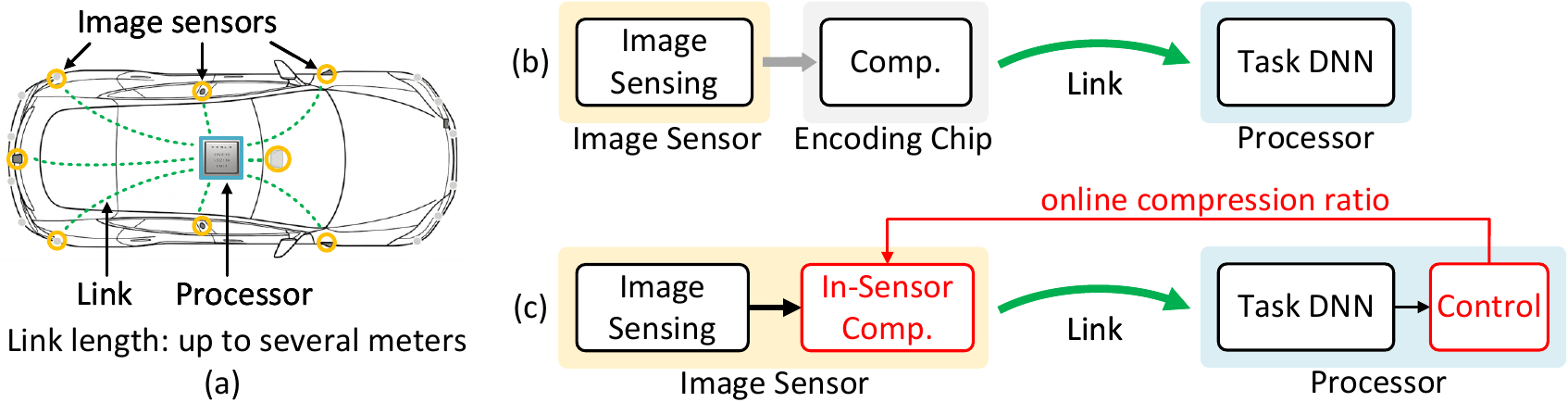} 
   \caption{(a) Multi-camera autonomous vehicle with lengthy data links. (b) Conventional vision system consists of an image sensor chip, a potential encoding chip, and a processor, which executes a task DNN. A large amount of data is sent from the sensor to the processor. (c) Proposed adaptive compression system, which dynamically adjusts the compression ratio based on the task DNN's performance.}
   \label{System_comparison}
\end{figure}

Computer vision systems are essential for autonomous vehicles to enable perception tasks such as detection \cite{ge2021yolox, wang2023yolov7}, tracking \cite{zhang2022bytetrack}, lane following \cite{tabelini2021keep}, drivable area segmentation \cite{Barua2019Navigation, janai2020computer, bojarski2016end}. 
Tesla’s Full Self-Driving (FSD) system is a representative example of the commercial adoption of such computer vision systems \cite{coffin2019building}. Its hardware platform employs eight image sensor chips (each with 1280$\times$960 resolution at 36 frames per second (fps) and 12-bit pixel depth) to achieve full 360-degree perception. 
These sensors are distributed around the vehicle perimeter, sending image streams to a centrally located processor, typically a system-on-chip (SoC) \cite{talpes2020compute, bannon2019computer}, which performs computer vision tasks using DNN models such as HydraNets \cite{mullapudi2018hydranets}.
Fig.~\ref{System_comparison}~(a) shows such a multi-camera automotive scenario. Fig.~\ref{System_comparison}~(b) shows the typical simplified pipeline of the vision system, which consists of a sensor, an encoding chip (e.g., H.264), and an SoC (or a processor). The communication distance between sensors and the processor can reach several meters.

One of the key challenges of the system lies in the massive data traffic over the long communication link. 
In Tesla's example, the link between distributed sensor chips and the central SoC must support data rates over 10~gigabits~per~second (Gbps). Such a large data rate not only results in a significant amount of energy but also requires highly complex link hardware \cite{damaj2022future}. 
As the industry projects to adopt more cameras, higher resolution, and growing frame rates, we expect this \textit{link} challenge to become even more severe.


Conventional compression methods, such as H.264 \cite{H264, H264_2}, can efficiently reduce data rates, but they are not appropriate in such real-time vision systems due to two key limitations. 
First, these algorithms have a high latency, as they exploit temporal redundancies among subsequent frames. They also have high computational complexity, increasing energy consumption and processing delay, making them unsuitable for real-time vision tasks and resource-constrained platforms.
Second, conventional compression frameworks typically operate with fixed user-defined settings, such as a target bitrate. However, real-world visual environments are inherently dynamic: scene content, motion patterns, and task difficulty vary over time. A fixed compression setting cannot adapt to these variations, resulting in either unnecessary bandwidth usage or task performance degradation.


To address these limitations, we propose OASIS, a closed-loop video compression framework that integrates lightweight in-sensor compression with task-aware adaptive control. As shown in Fig.~\ref{System_comparison}~(c), OASIS consists of three components: (1) an image sensor with configurable analog compression capability, (2) a downstream task DNN, and (3) a closed-loop feedback control. 
Unlike conventional complex codec-based solutions, OASIS employs a custom image sensor with an embedded analog compression processor, enabling low-complexity data reduction directly at the sensing front-end. Besides, the system dynamically adjusts the sensor’s compression ratio based on real-time task performance, enabling OASIS to maintain task accuracy while reducing data under changing scene conditions. Our main contributions are as follows:

\begin{itemize}
\item 
We propose a closed-loop feedback control to jointly optimize data reduction and task performance. It can adaptively modulate the real-time compression ratio for the next video frames based on the current task outputs.
\item 
We develop a hardware model for a 2.1-MP image sensor with a configurable analog compression processor. It predicts the sensor’s energy consumption, latency, and image quality under various compression ratios.
\item 
We validate the proposed system on multiple computer vision tasks, including multiple object tracking, traffic object detection, drivable area segmentation, and lane detection. It achieves an average 6$\times$ data compression and a 2.5$\times$ latency reduction, with a drop in task performance of less than 1.5\%.

\end{itemize} 

\section{Related Work}
\label{sec:related}

\subsection{Image and video compression} \label{background_image_compression}
Traditional image compression methods, such as JPEG, are designed to remove information that is irrelevant to the human eye. Standard video codecs, such as H.264 \cite{H264, H264_2} and H.265 \cite{H265}, also exploit temporal and spatial redundancies between frames to achieve a high compression ratio. Some prior works, however, have hypothesized that these methods may be unsuitable for DNN-based inference \cite{liu2018deepn}. In light of this, several works proposed to fine-tune compression algorithms to preserve features that are critical to DNN inference \cite{choi2020task, ye2023accelir, liu2025efficient, reddy2021pragmatic, xie2019source, xie2022bandwidth, JEONG20231, 10585292, shimauchi2004jpeg_resolution}. For example, Choi \etal \cite{choi2020task} proposed a DNN model to learn a JPEG quantization table optimized for a target task. Ye \etal \cite{ye2023accelir} trained a network to select the optimal compression ratios for each image block to reduce image restoration computation. Besides, Liu \etal \cite{liu2025efficient} introduced a method that leverages feature subset selection and task-specific tuning to optimize compression for both machine vision tasks and human perception. Other works \cite{xie2019source, xie2022bandwidth, JEONG20231, 10585292} further developed task-driven optimization frameworks that adjust color space, bitrate, or feature retention. However, all these fine-tuning works increase algorithmic complexity and overlook hardware and physical constraints, resulting in higher computation overhead and latency. Most of them also target the pre-deployment implementation, lacking the ability to handle dynamically changing scenes. 

\subsection{In-sensor compression hardware}  \label{background_in_sensor_compression}
Recent studies have explored in-sensor processors to reduce the output data size of sensor chips. Finateu \etal \cite{finateu20205} developed an event-based vision sensor chip with asynchronous sampling. Ma \etal \cite{ma2023leca} proposed an in-sensor analog processing technique, compressing raw images into low-dimensional, adaptively quantized features. Mukherjee \etal \cite{9462938} proposed to embed computing hardware in pixels to control pixel depth based on regions of interest. Ragusa \etal \cite{ragusa2024combining} proposed sensor hardware that adopts compressive sensing algorithms. However, their hardware are mostly hard-wired for specific applications and cannot support modern computer vision tasks.

On the other hand, for lower computation complexity, several recent works demonstrate image sensor chips (or sub-modules) with all-analog or analog-mixed-signal (AMS) transform-based compression hardware \cite{AJPEG, kumar202365, kawahito1997cmos}. In particular, AJPEG \cite{AJPEG} presents an end-to-end image sensor chip with an in-sensor analog DCT processor and 128-by-128 pixel arrays. It performs DCT on the raw pixel output, quantizes the DCT results, and digitizes only a few critical coefficients based on a user-defined target compression ratio. AJPEG can compress up to 10.7 times while preserving an acceptable output quality, with an average PSNR of 32 dB. It achieves an energy consumption of 26.4 pJ/pixel at 252 frames per second (fps). However, most prior works in this area lack exploration of integrating sensors into computer vision systems.

\section{Methodology}       \label{Methodology}
\subsection{System architecture}    \label{Methodology_system_architecture}

\begin{figure*}[tbp]
  \centering
  \includegraphics[width=0.98 \textwidth]{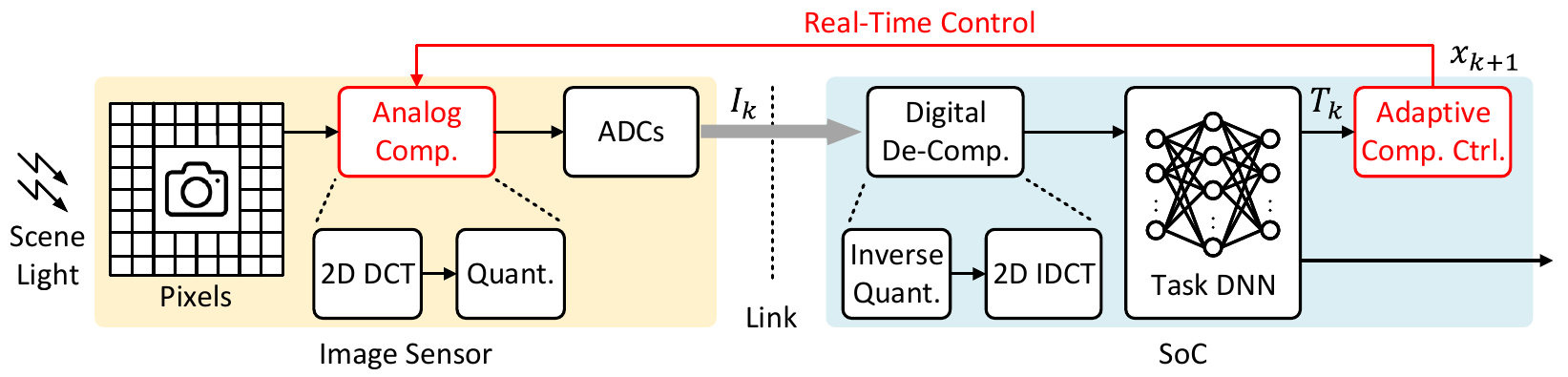}  
  \caption{ The proposed online adaptive video compression system. It consists of an image sensor, an analog compression processor, a task-specific DNN model, and a closed-loop control algorithm. It dynamically sets an optimal compression ratio for each frame.}
  \label{system_architecture}
\end{figure*}

Fig.~\ref{system_architecture} shows the architecture of the proposed online adaptive video compression system. It consists of an image sensor with an in-sensor compression processor, a task DNN, and a closed-loop control algorithm. The sensor performs compression on-chip. The closed-loop controller monitors both the real-time compression ratio and the corresponding DNN confidence outputs, then applies a gradient-descent-based algorithm to determine the optimal compression setting for the next frame. The following subsections provide a detailed description of the sensor model, task DNN, and control algorithm.

\subsection{Image sensor with in-sensor compression} \label{Methodology_sensor_model}

\begin{figure}[t]
    \centering 
    \includegraphics[width=1\linewidth]{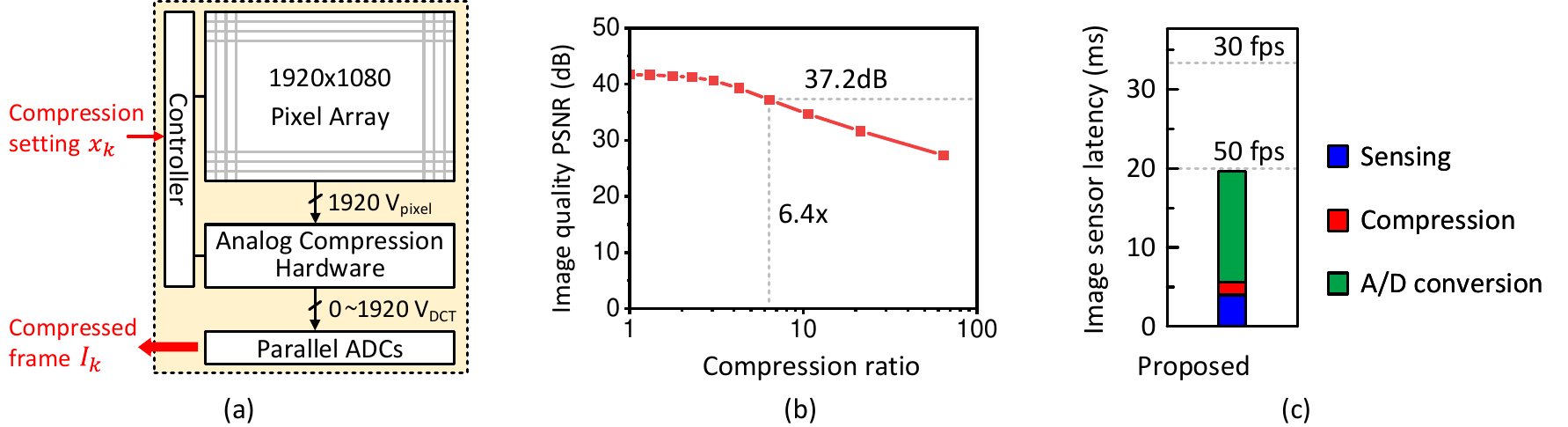}
    \caption{(a) The proposed 1080p image sensor with in-sensor analog compression hardware.
    (b) Image quality across compression ratios.
    (c) Sensor latency breakdown. }
    \label{Sensor_model}
\end{figure}

We construct an image sensor model based on AJPEG \cite{AJPEG}. It can capture an 8-bit 1920$\times$1080-pixel video stream at 30 fps, with a raw output data rate of 62.2 Mpix/s ($\sim$1.49 Gbps). Fig.~\ref{Sensor_model}(a) shows the block diagram of the proposed image sensor. It contains a 1920×1080 pixel array and an analog DCT processor with 240 DCT cores. Each core interfaces with eight columns of the pixel array. The DCT core is assumed to be identical to the one in \cite{AJPEG}. We estimate that the total active area of the image sensor is 3.3 mm$^2$ in a 0.18-$\mu$m CMOS, and the analog DCT processor occupies 17\% of the total area. 

The image sensor can compress the raw pixel output using the analog DCT processor. Specifically, each DCT core transforms 8$\times$8 pixel data from the spatial domain to the frequency domain, producing 8$\times$8 DCT coefficients. The low-frequency DCT coefficients contain the most relevant image information. Therefore, we compress the data by discarding high-frequency DCT coefficients. Since we have a total of 64 coefficients, we can discard none to all of them. The sensor has an input that sets the number of coefficients to discard, denoted as $x_k$, and as a result, the compression ratio is defined as: $CR=64/(64-x_k)$. 

We model the image quality as a function of compression ratios. Two factors affect image quality: the selective removal of DCT coefficients during the compression process and the variability of analog computing hardware. The former can be systematically captured during the inverse DCT (IDCT) process. To model the latter, we add 10\% random noise to the decompressed pixel values based on simulation results. Fig.~\ref{Sensor_model}~(b) shows the resulting image quality. It has a PSNR of 37.2 dB at 6.4$\times$ compression. 

We also evaluate the latency performance of the proposed sensor. For each row, the sensing time (capturing photons and producing current/voltage) is 4 ms; the in-sensor compression takes 1.5~$\mu$s and the analog-to-digital conversion (ADC) takes 13~$\mu$s \cite{AJPEG}. In the architecture, the sensing time is shared across all rows, whereas the subsequent compression and ADC operations are performed sequentially on a row-by-row basis. Based on this information, as shown in Fig.~\ref{Sensor_model}~(c), we can estimate the total per-frame latency to be about 19.6~ms for a 1080p sensor (1080 rows), which corresponds to a maximum frame rate of roughly 50 fps. Note that the compression process takes only 8\% of the total latency.

\subsection{Task performance}  \label{Task DNN}

\begin{figure}[t]
  \centering
   \includegraphics[width=0.45\linewidth]{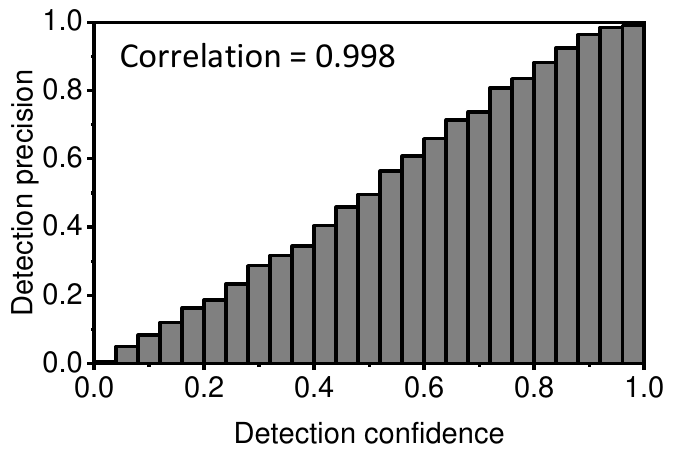}    
    \caption{Binned confidence values vs. detection precision ground truth within each bin, showing a strong positive correlation. The experiment considers the YOLO model and the COCO dataset.}
\label{detection_conf_vs_precision}
\end{figure} 

Our adaptive compression system supports various autonomous driving tasks, including multi-object tracking, traffic object detection, drivable area segmentation, and lane detection. These task-specific DNNs can be integrated into our system without architectural modifications or hyperparameter tuning. 

Ideally, the feedback controller should adjust the compression ratio based on task performance (i.e., accuracy). However, it requires ground-truth labels, which are unavailable post-deployment. Therefore, we use the confidence score produced by the task DNN as a proxy for inference accuracy. We validate this using the YOLO detector on the COCO dataset. Fig.~\ref{detection_conf_vs_precision} shows the relationship between the predicted confidence score and the corresponding ground-truth precision. The strong positive correlation (correlation coefficient = 0.998) confirms that the confidence score is a reliable surrogate for accuracy for the purpose of control.

\subsection{Online adaptive control algorithm} \label{Methodology_control_algorithm}

We develop the online adaptive controller algorithm to jointly optimize data size and task performance. 
Algorithm~\ref{Algorithm1} summarizes the proposed algorithm, which operates as follows. First, it receives a real-time compressed video frame $\mathcal{I}_k$ from the image sensor and decompresses it through IDCT (line 3). Then it performs the task DNN (e.g., object detection and tracking), which produces the output tensor ${T}_k$ at time index $k$ (line 4). The output tensor consists of $N_k$ pairs of objects $\mathbf{id}_{k, i}$ and confidence scores $\mathbf{conf}_{k, i}$.
Then, the control algorithm updates the objective function $\mathcal{J}_k$ (line 5), calculates its gradient (line 6), and updates the compression ratio setting $x_{k+1}$ with the updated step size $\eta_k$ (lines 7 and 8). 

The objective function is designed to jointly optimize data reduction and task performance. It is defined as 
\begin{equation}
    \mathcal{J}_k =   64^{(1-\alpha)}\cdot x_k^\alpha \cdot \overline{  \mathbf{conf}_k} , 
\label{objective_equation}
\end{equation}
where $x_k$ denotes the number of DCT coefficients discarded during compression, serving as the compression setting input to the image sensor (see Sec.~\ref{Methodology_sensor_model}). $\overline{\mathbf{conf}_k}$ represents the average confidence score computed over $N_k$ detected objects. $\alpha$ is a hyperparameter in [0,1]. It affects the stationary point of $\mathcal{J}_k$~\cite{yang2020nature} (i.e., the ideal optimal compression configuration):
\begin{equation}
    x_{sta} = \frac{\alpha \cdot \overline{\mathbf{conf}_k} }{- \partial\overline{\mathbf{conf}_k} /{\partial x} }, 
\label{optimal_compression_ratio}
\end{equation}
where a larger $\alpha$ favors more compression over task performance, while a smaller $\alpha$ tries to retain task performance. 

Note that it is not possible to directly differentiate $\mathcal{J}_k$ since $x_k$ is discrete and $\overline{\mathbf{conf}_k}$ is the task DNN output. Therefore, we approximate it with the numerical gradient as follows:
\begin{equation}
 \nabla\mathcal{J}_k = 
        \begin{cases}
            \frac{\mathcal{J}_k - \mathcal{J}_{k-1}}{x_k-x_{k-1}}, & \text{if } x_{k} \neq x_{k-1}    \\  
            \operatorname{sign}(\Delta_\mathbf{c})\cdot \mathbf{1}(|\Delta_\mathbf{c}|>\epsilon),  & \text{otherwise} 
        \end{cases}
\end{equation}

Also, recall that before updating the compression setting $x_{k+1}$, the algorithm updates the step size $\eta_k$. This is to balance control stability and response time. The step size is formulated as: 
\begin{equation}
\eta_k =\eta_{0} \cdot \frac{ 1 }{1 + \beta \cdot x_k},
\label{equation_beta}
\end{equation}
where $\beta$ is a hyperparameter in [0,1]. If $\beta \cdot x_k$ is large, it decreases the step size to slowly change the compression setting, making the feedback loop stable. On the other hand, if $\beta \cdot x_k$ is small, it changes the compression setting more aggressively, improving the response time.  

\begin{algorithm}[tb]
\caption{Task-based feedback control to modulate the compression ratio}  
\label{Algorithm1}  
\begin{algorithmic}[1] 
\Require Real-time video frame sequence $\left\{ \mathcal{I}_k \right\}_{k=1}^T$ 
   \Statex      Initial step size $\eta_{0}$ 
\Ensure Task DNN output $\left\{ \mathcal{T}_k \right\}_{k=1}^T$ 
   \Statex    Compression ratios $\left\{ {c}_k \right\}_{k=1}^T$, $c_k = \frac{64}{64-x_k} $   
\State Initialize compression setting $x_{1} \gets 0$
\For{$k \gets 1$ to $N$}    
    \State  Decompress the k-th frame $\mathcal{I}_k$ with $x_k$  
    \State  Apply task DNN, $\mathcal{T}_k = \left\{ \mathbf{id}_{k, i}, \mathbf{conf}_{k,i} \right\}_{i=1}^{N_k} $
    \State Update objective function $\mathcal{J}_k $           
    \State Update gradient $\nabla\mathcal{J}_k$ 
    \State Update step size $\eta_k   $ 
    \State Update compression setting for $\mathcal{I}_{k+1}$: $ x_{k+1} = x_k + \operatorname{round} ( \eta_k \cdot \nabla\mathcal{J}_k $ )      
\EndFor
\end{algorithmic} 
\end{algorithm}

\section{Experiments} \label{Experiments}

\subsection{Setting} \label{Experiments_settings}

\paragraph{Task DNN model.}
We validate the proposed video compression system with multiple vision tasks, including multiple object tracking, traffic object detection, drivable area segmentation, and lane detection.
For the multiple object tracking task, we implement a two-stage DNN based on ByteTrack \cite{zhang2022bytetrack}. ByteTrack primarily consists of YOLOX \cite{ge2021yolox}, which serves as the backbone for object detection, and the BYTE algorithm \cite{zhang2022bytetrack}, which utilizes the detection results from YOLOX for tracking.
In addition, for traffic object detection, drivable area segmentation, and lane detection, we implement a DNN model based on YOLOP \cite{wu2022yolop}.

\paragraph{Datasets.}
We evaluate our adaptive video compression system on MOT17 \cite{milan2016mot16}, MOT20 \cite{dendorfer2020mot20},  DanceTrack \cite{sun2022dancetrack} and BDD100K \cite{yu2018bdd100k} datasets.
MOT17 comprises 14 video sequences: seven for training and seven for testing. Most sequences have a resolution of 1920$\times$1080 at 30 frames per second (fps), captured using either static or moving cameras. We use the first half of every video in the training set for training, and the last half for validation and ablation studies.  
MOT20 is designed for highly crowded pedestrian detection, with an average of more than 150 people per frame. It contains eight sequences (four for training and four for testing) at a frame rate of 25 frames per second.
DanceTrack focuses on multi-human tracking with similar appearances but diverse motions and articulations. It contains 100 sequences: 40 for training, 25 for validation, and 35 for testing.
BDD100K is a large-scale driving dataset with 100k annotated images and 10k video sequences, covering diverse driving scenarios and weather conditions with annotations for 10 tasks. It is divided into 70k training images, 10k validation images, and 20k testing images.

\paragraph{Implementation details.}
For the multiple object tracking task, we perform a detection stage followed by a tracking stage. The detection stage is based on YOLOX.  For MOT17, we trained YOLOX using a combination of MOT17, CrowdHuman \cite{shao2018crowdhuman}, Cityperson \cite{zhang2017citypersons}, and ETHZ \cite{ess2008mobile} datasets. For MOT20, we trained YOLOX using MOT20 and CrowdHuman datasets. For DanceTrack, we trained YOLOX with only DanceTrack's training dataset. The tracking stage is based on the BYTE algorithm. We use the same settings as in \cite{zhang2022bytetrack}. 
For the multi-task problem, we train the YOLOP model end-to-end on BDD100K and jointly evaluate three tasks, following the practices in \cite{wu2022yolop}. As test annotations are unavailable, we perform evaluations on the BDD100K validation set.
All experiments are performed on a server equipped with dual Intel Xeon E5-2690 processors and an NVIDIA Titan RTX GPU.

\paragraph{Matrics.}
Task performance is evaluated using the following metrics. 
For multiple object tracking, we report MOTA, IDF1, and HOTA. MOTA focuses on detection accuracy, IDF1 emphasizes identity preservation, and HOTA balances detection accuracy and overall tracking quality.
For object detection, we use Recall (Rcll), Precision (Prcn), and mAP@50. Recall measures the proportion of ground truth objects correctly detected, while precision measures the correctness of detected objects. mAP@50 is the mean average precision at an intersection-over-union (IoU) threshold of 0.5, capturing both localization and classification performance. 
For drivable area segmentation and lane detection, we evaluate accuracy and mean intersection over union (mIoU), where accuracy is the overall pixel-wise accuracy and mIoU is the average overlap between the predicted and ground-truth regions.
Additionally, we define CR as the average compression ratio across all video frames. Link latency includes the latencies of compression, data transmission, and decompression. 
In the tables, \textbf{bold} and \underline{underline} indicate the best and the second-best results, respectively.
 
\subsection{Experiment results} \label{Experiments_performance}

\begin{figure}[t]
    \centering 
    \includegraphics[width=0.8\linewidth]{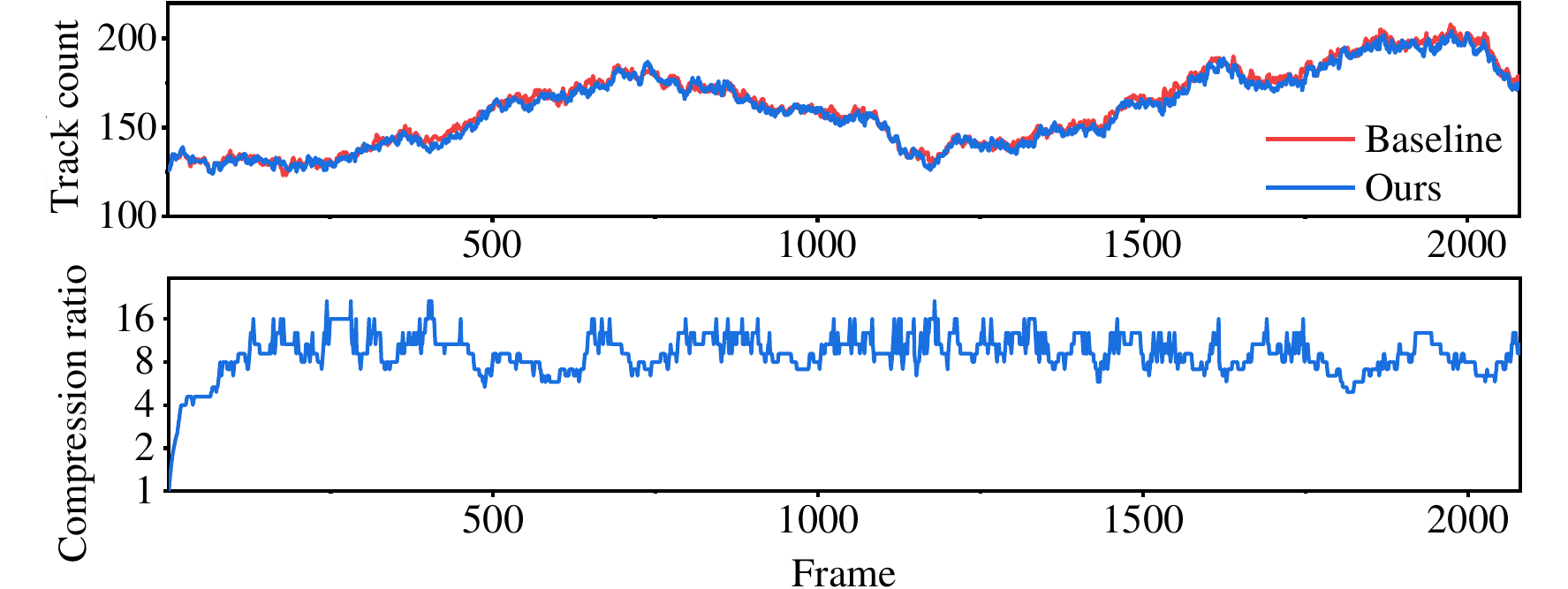} 
    \caption{(a) Tracking performance of the no-compression baseline \cite{zhang2022bytetrack} (red) and our proposed adaptive compression (blue). They are closely matched. The MOT20-04 is used for the experiment. (b) The controller automatically determines the compression ratios on a frame-by-frame basis.} 
  \label{object_tracking}
\end{figure} 

\paragraph{Online compression and tracking.} 
Fig.~\ref{object_tracking} shows the example outputs of the proposed adaptive compression system on the MOT20-04 sequence. The scene features a crowded public square at night, where tracking is challenging due to object movements, heavy occlusion, and lighting variations across frames. However, our adaptive compression system successfully adjusts the compression ratio in response to real-time tracking metrics. The top of the figure shows tracking counts across frames, where our system closely follows the baseline. The bottom of the figure shows the compression ratios across frames, which vary from 1$\times$ to 21$\times$ with an average ratio of 8.6$\times$.

\paragraph{Link power consumption.}  
We evaluate the cost of the link in the proposed adaptive compression system. The cost includes the power consumption and latency of three components: compression, data transmission, and decompression. We do not include the cost of other components, as they remain the same as baseline systems (e.g., task DNN) or add a negligible amount of cost (e.g., the controller).

Compression and decompression add additional power consumption, while reducing data size helps save the power consumption associated with data transmission. We model the link power $P_L$ as follows: 
\begin{equation}
P_{L} = P_{c} +P_{t}+P_{dc},
\end{equation}
where $P_c$ is the power consumption for compression, $P_t$ is that for data transmission, and $P_{dc}$ is that for decompression. We consider the 1080p 30-fps video input. 

We estimate the compression power: $P_c=0.34$~mW, based on the energy consumption measurement of 5.5~pJ/pixel \cite{AJPEG}. We estimate the data transmission power consumption using $P_t = {192.5}/c$~mW, where $c$ is the compression ratio. 
It is based on the 4-Gbps FPD-Link III circuit used in Tesla vehicles, with an energy consumption of 129 pJ/bit, as reported in prior studies \cite{DS90UB953, DS90UB954}. 
Finally, we estimate the power consumption for decompression as $P_{dc} = 0.97$ mW. The decompression involves mostly IDCT operations. A fast 8$\times$8 IDCT algorithm requires roughly 600 operations per block \cite{arai1988fast,feig1992fast}. To serve the target 1080p 30-fps video, the processor should achieve a computing throughput of approximately 1.75 giga operations per second (GOPS). We assume the energy efficiency of the processor to be 1.8~TOPS/W, based on \cite{talpes2020compute,bannon2019computer}.

\begin{figure}[t]
    \centering 
    \includegraphics[width=0.8\linewidth]{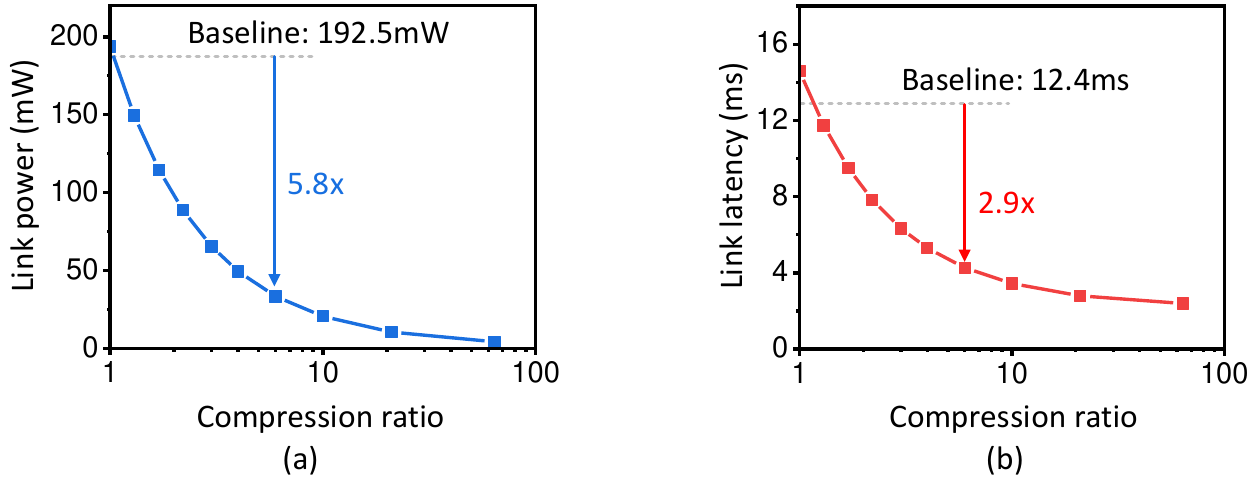}
    \caption{(a) Communication power consumption across compression ratios.
    (b) Communication latency across compression ratios. The baseline performs no compression.}
    \label{Energy_latency}
\end{figure} 

Combining the above estimates, we formulate the power consumption for the link as $P_L = 1.31 + {192.5}/{c}$ mW. Fig.~\ref{Energy_latency}(a) plots it over compression ratios. At a 6$\times$ compression ratio, the proposed systems can consume 5.8$\times$ less power in the link compared to the non-compression baseline. Note that the baseline only includes the power consumption for data transmission. The plot shows that the reduction in $P_{t}$ easily offsets the increase in $P_{c}$ and $P_{dc}$, making it particularly advantageous if the link length is long. 

\paragraph{Link latency.}
We formulate the link latency as follows: 
\begin{equation}
T_{L} = T_{c} + T_{t} + T_{dc},
\end{equation}
where $T_{c}$, $T_{t}$, and $T_{dc}$ denote the latencies for compression, data transmission, and decompression, respectively. Again, considering the 1080p 30-fps video, we estimate the compression latency to be $T_{c} \approx 1.6$~ms per frame. This is based on the measurement results in \cite{AJPEG}, where an analog DCT processor takes 1.5~$\mu$s to process one row of the input data on average. 
Also, we estimate the latency of data transmission to be $T_{t} \approx 12.4/c$~ms based on the 4-Gbps FPD-Link III circuit data rate \cite{DS90UB953, DS90UB954}. 
Finally, we estimate the latency of decompression to be approximately $T_{dc} \approx 0.6$ ms. Performing IDCTs for a 1080p frame involves roughly 58 million operations \cite{arai1988fast,feig1992fast}. Some of the recent SoCs report a throughput of over 70 TOPS, which could complete the decompression in 1 $\mu$s; but we consider an NVIDIA Titan RTX GPU (released in 2018) and conservatively estimate the latency. 
Combining the above estimates, we predict the link latency as $T_L = 2.2 + 12.4/c$~ms. Fig.~\ref{Energy_latency}(b) plots this equation over compression ratios. At a $6\times$ compression ratio, the latency is approximately 4.3 ms, showing a 2.9$\times$ reduction as compared to the no-compression baseline.



\paragraph{Benchmark evaluation.}

\begin{table}[tb]  
\caption{Performance comparison of the no-compression baseline \cite{zhang2022bytetrack}, H.264 \cite{H264,H264_2}, and the proposed adaptive compression system on the MOT17 validation dataset \cite{milan2016mot16}. Our method preserve high accuracy while reducing link latency by 2.75$\times$ and data size by 5.5$\times$ on average. Note that we evaluate a range of compression ratios in H.264, as the optimal ratio depends on scene characteristics and cannot be predetermined.}
\label{SOTA_Comparison}
\centering
\setlength{\tabcolsep}{0.5mm}
\resizebox{\linewidth}{!}{
\begin{tabular}{c|cc|cc|ccc|cccc}
\hline
\multirow{2}{*}{Method} & \multicolumn{2}{c|}{Compression} & \multicolumn{2}{c|}{Detection (\%) ↑} & \multicolumn{3}{c|}{Tracking   (\%) ↑} & \multicolumn{4}{c}{Link latency (ms)↓} \\
                        & Type                     & Avg CR             & Rcll          & Prcn          & MOTA          & IDF1          & HOTA          & Comp   & Trans   & De-comp  & Sum   \\ \hline
Baseline                & No                       & 1$\times$          & \underline{72.9}  & 80.8              & \textbf{77.5}     & \textbf{79.7}     & \textbf{67.5}     & \textbf{0.0} & 12.4           & \textbf{0.0}  & \underline{12.4}  \\ \hline
\multirow{4}{*}{H.264}  & \multirow{4}{*}{Fixed}   & 2$\times$          & \textbf{73.0}     & 80.9              & \underline{77.4}  & 77.9              & \underline{66.5}  & 7.1           & 6.2            & 12.0          & 25.4  \\
                        &                          & 5.5$\times$        & 71.7              & 80.8              & 76.0              & 77.3              & 66.1              & 5.4           & 2.3            & 10.0          & 17.7  \\
                        &                          & 10$\times$         & 69.5              & \underline{81.0}  & 74.5              & 75.2              & 64.1              & 5.1           & \underline{1.2}            & 9.1           & 15.4  \\
                        &                          &  20$\times$        & 63.9              & 80.8              & 68.8              & 72.0              & 61.2              & 5.0           & \textbf{0.6}   & 8.0           & 13.6  \\ \hline
Ours                    & Adaptive                 & 5.5$\times$        & 70.8              & \textbf{81.2}     & 75.9              & \underline{78.0}  & 66.2              & \underline{1.6} & 2.3            & \underline{0.6}           & \textbf{4.5}   \\ \hline
\end{tabular}}
\end{table}

\begin{table}[tb]
\centering
\caption{Performance comparison of the no-compression baseline \cite{zhang2022bytetrack} and our proposed adaptive compression system. We use MOT17 \cite{milan2016mot16}, MOT20 \cite{dendorfer2020mot20}, and DanceTrack \cite{sun2022dancetrack} test sets. Our adaptive system reduces the data size by 4.8-6.4$\times$ and the link latency by 2.4-3.0$\times$ on average, while retaining the accuracy performance close to the baseline.}
\label{MOT_Comparison}
\setlength{\tabcolsep}{0.7mm}
\resizebox{\linewidth}{!}{
\begin{tabular}{c c|cc|cc|ccc|c}
\hline
\multirow{2}{*}{Dataset}    & \multirow{2}{*}{Method} & \multicolumn{2}{c|}{Compression} & \multicolumn{2}{c|}{Detection   (\%) ↑} & \multicolumn{3}{c|}{Tracking (\%) ↑} & {Latency (ms)↓}    \\
                            &          & Type      & Avg CR                & Rcll          & Prcn          & MOTA          & IDF1          & HOTA          & Sum\\ \hline
\multirow{2}{*}{MOT17}      & Baseline & No        & 1$\times$           & \textbf{84.3} & 94.4          & \textbf{78.8} & \textbf{75.9} & \textbf{62.4} & 10.6 \\
                            & Ours     & Adaptive  & \textbf{4.8$\times$}& 82.5          & \textbf{95.1} & 77.8          & 74.7          & 61.5          & \textbf{4.4} \\ \hline
\multirow{2}{*}{MOT20}      & Baseline & No        & 1$\times$           & \textbf{78.5} & 95.5          & \textbf{74.6} & \textbf{75}   & \textbf{60.8} & 12.4 \\
                            & Ours     & Adaptive  & \textbf{5.3$\times$}& 77.8          & \textbf{95.6} & 74            & 74.4          & 60.4          & \textbf{4.5} \\ \hline
\multirow{2}{*}{DanceTrack} & Baseline & No        & 1$\times$           & \textbf{77}   & \textbf{81.1} & \textbf{88.3} & \textbf{51.3} & \textbf{46.1} & 12.4 \\
                            & Ours     & Adaptive  & \textbf{6.4$\times$}& 75.9          & 81            & 87.6          & 50.2          & 45.1          & \textbf{4.1} \\ \hline
\end{tabular}}
\end{table}

Table~\ref{SOTA_Comparison} compares the proposed adaptive compression system with the uncompressed baseline and the state-of-the-art H.264 codec on the MOT17 dataset. Our system automatically adjusts the compression ratio according to task performance. On the other hand, H.264 requires a pre-defined compression ratio and cannot adapt dynamically. It is challenging to find a single setting that performs well under all conditions, including those that are unknown. For this reason, using a fixed compression ratio generally requires a large safety margin. 
As shown in Table~\ref{SOTA_Comparison}, different fixed CRs in H.264 lead to significant variations in detection and tracking results, indicating the difficulty of selecting an optimal compression level in advance. In contrast, our adaptive control system achieves a balanced point between data reduction and task performance, realizing a 5.5$\times$ compression with less than 1.5\% accuracy degradation. Furthermore, it achieves a 2.75$\times$ reduction in link latency. 
 
Table~\ref{MOT_Comparison} presents the results across multiple benchmarks, including MOT17, MOT20, and DanceTrack. Our adaptive compression system maintains consistent accuracy across a diverse range of tasks, achieving data reductions of 4.8-6.4$\times$. We observe small degradations in Rcll, MOTA, IDF1, and HOTA. They are primarily attributed to missed detections under high compression ratios. The link latency is reduced by 2.4-3$\times$ across all datasets. The results confirm that the proposed method not only preserves tracking accuracy under substantial compression but also enables real-time efficiency, making it suitable for edge deployment.
   
\paragraph{Model size.}

\begin{table}[t]
\centering
\caption{Impact of the task DNN model size. YOLOX-X has 99M parameters, while YOLOX-S has only 9M parameters. Our system achieves a similar improvement when applied to both models.}
\label{MOT17_ablation}
\setlength{\tabcolsep}{1mm}
\resizebox{0.57\linewidth}{!}{
\begin{tabular}{c|cc|ccc} \hline
Backbone                 & Method   & Avg CR & MOTA  & IDF1  & HOTA  \\ \hline
\multirow{2}{*}{YOLOX-X} & Baseline & 1$\times$   & \textbf{77.5} & \textbf{79.7} & \textbf{67.5} \\ 
                         & Ours     & \textbf{5.3$\times$} & 76.1 & 77.3 & 65.9 \\ \hline
\multirow{2}{*}{YOLOX-S} & Baseline & 1$\times$   & 69.4 & 70.7 & \textbf{59.5} \\
                         & Ours     & \textbf{5.1$\times$} & \textbf{69.5} & \textbf{70.9} & 58.8 \\ \hline
\end{tabular}}
\end{table}

The previous benchmark evaluation was based on YOLOX-X, which contains 99M parameters. To verify the robustness of our system with a smaller task DNN, we conduct similar experiments using YOLOX-S, which has 11 times fewer parameters than YOLOX-X. The results in Table~\ref{MOT17_ablation} indicate that, when applied to the smaller model, our compression system has minimal impact on task performance. The average compression ratio is also similar. In some metrics (MOTA and IDF1), the performance even slightly exceeds that of the uncompressed case, showing that the proposed closed-loop control performs equally well regardless of the model size.


\paragraph{Robustness.}
We conduct experiments to evaluate the robustness of the proposed closed-loop control system. Following the practices in \cite{boncelet2009image, boyat2015review}, we add three types of noise to the video input of the system. First, we add a fixed/static noise to all pixels in all frames. The magnitude of the noise is fixed to 15. Note that each pixel is 8 bits, ranging from 0 to 255. It emulates a steady sensor offset and an almost static illumination change (e.g., overexposed scenes or color temperature drift). Second, we add temporal noise to all pixels of 5\% of the frames randomly selected. The magnitude of the temporal noise follows a uniform distribution between ±30. It simulates time-varying disturbances, such as random flicker and transient interference. Finally, we add random noise to all pixels in all frames. The magnitude follows a Gaussian distribution with a mean of 0 and a standard deviation ($\sigma$) of 10. It models noise from pixel circuits and sensor readout circuits. Table~\ref{Noise_Performance} summarizes the experiment's results. The proposed control system maintains consistent task performance and compression ratios across various combinations of the noises, confirming the robustness of the proposed control system. 


\begin{table}[tb]
\centering
\caption{Impacts of the fixed, random, and temporal noise on the proposed adaptive compression system. In the worst-case noise condition, it retains all accuracy performance with less than 1.2\% degradation. 
The MOT17 dataset is used for the experiment \cite{milan2016mot16}.}
\label{Noise_Performance}
\setlength{\tabcolsep}{0.5mm}
\resizebox{0.9\linewidth}{!}{
\begin{tabular}{c|ccc|c|cc|ccc}
\hline
\multirow{2}{*}{Method} & \multicolumn{3}{c|}{Noise}                                          & \multicolumn{1}{c|}{Compression} & \multicolumn{2}{c|}{Detection   (\%) ↑} & \multicolumn{3}{c}{Tracking (\%) ↑} \\
                        & Fixed                & Temporal               & Random              & Avg CR        & Rcll               & Prcn              & MOTA       & IDF1      & HOTA      \\ \hline
\multirow{5}{*}{Ours}   & $\times $     & $\times $     & $\times $     & \textbf{5.5} & \textbf{70.8} & 81.2          & 75.9          & \textbf{78.0}  & \textbf{66.2} \\
                        & $\checkmark$  & $\times $     & $\times $     & 5.3          & 70.4          & \textbf{81.5} & \textbf{76.0} & 76.6           & 65.5          \\
                        & $\times $     & $\checkmark$  & $\times $     & 5.4          & 70.6          & 81.3          & 75.8          & 77.6           & 65.7          \\
                        & $\times $     & $\times$      & $\checkmark$  & 5.4          & 69.7          & \textbf{81.5} & 75.7          & 77.4           & 65.3          \\
                        & $\checkmark$  & $\checkmark$  & $\checkmark$  & \textbf{5.5} & 69.6          & 81.4          & 75.2          & 77.0           & 65.2          \\ \hline
\end{tabular}}
\end{table}

\begin{table}[t]
\centering
\caption{The proposed adaptive compression system is verified across multiple tasks, including object detection, area segmentation, and lane detection. The proposed control retains the accuracy performances and the average compression ratio across the tasks. The baseline employs no compression \cite{wu2022yolop}. The BDD100K dataset is considered \cite{yu2018bdd100k}.}
\label{Multitask_Performance}
\setlength{\tabcolsep}{0.8mm}
\resizebox{\linewidth}{!}{
\begin{tabular}{c|cc|cc|cc|cc}
\hline
\multirow{2}{*}{Methold} & \multicolumn{2}{c|}{Compression} & \multicolumn{2}{c|}{Traffic   object detection} & \multicolumn{2}{c|}{Drivable   area seg.} & \multicolumn{2}{c}{Lane   detection} \\
                         & Type              & Avg CR         & mAP50(\%)              & Recall(\%)             & Acc.(\%)              & mIOU(\%)              & Acc.(\%)        & mIOU(\%)       \\ \hline
Baseline                 & No                & 1$\times$            & \textbf{76.5}  & \textbf{89.2} & \textbf{97.4}  & \textbf{91.5} & \textbf{70.5} & \textbf{62.3} \\
Ours                     & Adaptive          & \textbf{5.3$\times$} & 76             & 88.8          & \textbf{97.4}  & \textbf{91.5} & 70.1          & \textbf{62.3} \\ \hline
\end{tabular}}
\end{table}

\paragraph{Scalability.}
We conduct experiments to evaluate the scalability of our control system to other perception tasks. To achieve this, we apply the proposed control system to three additional tasks: traffic object detection, drivable area segmentation, and lane detection. We use the YOLOP model and the BDD100K dataset. Specifically, the proposed controller takes the combined confidences of all three tasks and jointly optimizes data reduction and multi-task accuracy. Table~\ref{Multitask_Performance} summarizes the experimental results. The proposed controller achieves a 5.3$\times$ compression ratio on average, while preserving nearly the same performance across all tasks. These results confirm that the proposed framework generalizes effectively across multi-task scenarios and scales reliably to broader vision applications.

\paragraph{Trade-off between compression and accuracy.}

We investigate the impact of the stationary point in the proposed control system, particularly the hyperparameter $\alpha$. As shown in (\ref{objective_equation}) and (\ref{optimal_compression_ratio}), $\alpha$ modulates the stationary compression setting ($x_{sta}$). We set $\alpha$ to three different values and evaluate the control system in terms of the average compression ratio and accuracy. Table~\ref{table_tracking_vs_cr} shows the results. When $\alpha$ is set to 1, the proposed system achieves comparable accuracy at a 6.4$\times$ compression ratio. On the other hand, when we set $\alpha$ to 0.5, it starts to outperform the baseline in the accuracy metrics at a lower compression ratio of 3.1$\times$. Since the original video contains high-frequency noise, mild compression removes some of this noise and improves task performance.

\begin{table}[t]
\centering
\caption{The impact of parameter settings in the adaptive compression system. By increasing the hyperparameter $\alpha$ in (\ref{objective_equation}), we can make the control system adjust the compression ratio more aggressively. But it may degrade the accuracy performance if the selected compression ratio is too high. We set $\alpha$ to balance the compression ratio and accuracy performance. The DanceTrack dataset is considered \cite{sun2022dancetrack}.}
\label{table_tracking_vs_cr}
\setlength{\tabcolsep}{0.8mm}
\resizebox{0.5\linewidth}{!}{
\begin{tabular}{c|cc|ccc} \hline
Method                  & $\alpha$           & Avg CR & MOTA & IDF1 & HOTA \\ \hline
Baseline                & N/A               & 1$\times$            & \textbf{88.3}     & 51.3              & 46.1  \\ \hline
\multirow{3}{*}{Ours}   & 1             & \textbf{6.4$\times$} & 87.6              & 50.2              & 45.1  \\
                        & 0.8            & 5.5$\times$          & 87.9              & 50.9              & 46.2  \\
                        & 0.5             & 3.1$\times$          & \textbf{88.3}     & \textbf{51.7}     & \textbf{46.4}  \\  \hline
\end{tabular}}
\end{table}

\paragraph{Controller update interval.}

\begin{table}[tb]
\centering
\caption{The proposed adaptive system can update the compression ratio every $N_f$ frames. A larger $N_f$ can relax the latency requirement of the feedback loop, offering a longer timing slack for a task DNN. However, it tends to degrade the compression ratio and the accuracy performance.}
\label{Update_Interval}
\setlength{\tabcolsep}{0.8mm}
\resizebox{0.65\linewidth}{!}{
\begin{tabular}{c|c|c|c|ccc} \hline
Method                  & $N_f$  & Slack (ms)    & Avg CR & MOTA & IDF1 & HOTA \\ \hline
Baseline                & N/A       & N/A           & 1$\times$             & \textbf{77.5}	    & \textbf{79.7}      & \textbf{67.5}         \\ \hline
\multirow{3}{*}{Ours}   & 1         & 13.7          & \textbf{5.5$\times$}  & 75.9	            & \underline{78.0}   & \underline{66.2}	      \\
                        & 2         & 47            & 5.1$\times$           & 76.0	            & 76.6               & 64.9	                  \\
                        & 3         & 80.3          & 5.1$\times$           & \underline{76.6}  & 76.1               & 65.1                     \\
                        & 4         & 113.6         & 3.5$\times$           & 76.4              & 77.3               & 65.8                    \\ \hline
\end{tabular}}
\end{table}

So far, we have configured our control system to update the compression ratio every frame. However, this is feasible only if the overall loop latency is less than the frame time (e.g., 33.3~ms for 30~fps). In practice, certain task DNNs may take a long time to complete an inference operation, and if so, the adaptive controller cannot update the compression ratio every frame. Therefore, to accommodate such cases, we conduct a set of experiments in which we configure the controller to update the compression ratio every $N_f$ frames. Slowly updating the compression ratio offers a longer timing slack for a task DNN to finish the inference task. Table~\ref{Update_Interval} summarizes the results. If $N_f$ is set to 1, the proposed system achieves the highest accuracy and compression ratio. As we set $N_f$ to a larger value, we observe that the system achieves slightly lower 
performance, particularly in IDF1 and HOTA, while maintaining the compression ratio. However, at $N_f$=4, the compression ratio starts to decrease, as the controller prioritizes maintaining accuracy over data reduction. 

\section{Conclusion} \label{conclusion}

In an autonomous vehicle, the massive data traffic between sensors and processors presents a critical challenge. To address this challenge, we propose online adaptive closed-loop video compression systems that dynamically adjust the compression ratio to jointly optimize data reduction and task performance. 
Validated across multiple tasks, the proposed adaptive compression system achieves an average of 6 times data reduction, 5.8 times lower power consumption, and 2.5 times shorter link latency, while maintaining task accuracy with less than 1.5\% degradation.  The experimental results also validate the robustness and scalability of the proposed adaptive system.


%
%
\bibliographystyle{splncs04}
\bibliography{main}

@String(ECCV  = {Eur. Conf. Comput. Vis.})

@String(ECCV  = {ECCV})

@article{janai2020computer,
  title={Computer vision for autonomous vehicles: Problems, datasets and state of the art},
  author={Janai, Joel and G{\"u}ney, Fatma and Behl, Aseem and Geiger, Andreas and others},
  journal={Foundations and trends{\textregistered} in computer graphics and vision},
  volume={12},
  number={1--3},
  pages={1--308},
  year={2020},
  publisher={Now Publishers, Inc.}
}

@article{talpes2020compute,
  title={Compute solution for tesla's full self-driving computer},
  author={Talpes, Emil and Sarma, Debjit Das and Venkataramanan, Ganesh and Bannon, Peter and McGee, Bill and Floering, Benjamin and Jalote, Ankit and Hsiong, Christopher and Arora, Sahil and Gorti, Atchyuth and others},
  journal={IEEE Micro},
  volume={40},
  number={2},
  pages={25--35},
  year={2020},
  publisher={IEEE}
}

@article{ge2021yolox,
  title={Yolox: Exceeding yolo series in 2021},
  author={Ge, Zheng and Liu, Songtao and Wang, Feng and Li, Zeming and Sun, Jian},
  journal={arXiv preprint arXiv:2107.08430},
  year={2021}
}

@inproceedings{zhang2022bytetrack,
  title={ByteTrack: Multi-Object Tracking by Associating Every Detection Box},
  author={Zhang, Yifu and Sun, Peize and Jiang, Yi and Yu, Dongdong and Weng, Fucheng and Yuan, Zehuan and Luo, Ping and Liu, Wenyu and Wang, Xinggang},
  booktitle={Proceedings of the European Conference on Computer Vision (ECCV)},
  year={2022}
}

@article{milan2016mot16,
  title={MOT16: A benchmark for multi-object tracking},
  author={Milan, Anton and Leal-Taix{\'e}, Laura and Reid, Ian and Roth, Stefan and Schindler, Konrad},
  journal={arXiv preprint arXiv:1603.00831},
  year={2016}
}

@article{dendorfer2020mot20,
  title={Mot20: A benchmark for multi object tracking in crowded scenes},
  author={Dendorfer, Patrick and Rezatofighi, Hamid and Milan, Anton and Shi, Javen and Cremers, Daniel and Reid, Ian and Roth, Stefan and Schindler, Konrad and Leal-Taix{\'e}, Laura},
  journal={arXiv preprint arXiv:2003.09003},
  year={2020}
}

@inproceedings{sun2022dancetrack,
  title={Dancetrack: Multi-object tracking in uniform appearance and diverse motion},
  author={Sun, Peize and Cao, Jinkun and Jiang, Yi and Yuan, Zehuan and Bai, Song and Kitani, Kris and Luo, Ping},
  booktitle={Proceedings of the IEEE/CVF conference on computer vision and pattern recognition},
  pages={20993--21002},
  year={2022}
}

@ARTICLE{H264,
  author={Wiegand, T. and Sullivan, G.J. and Bjontegaard, G. and Luthra, A.},
  journal={IEEE Transactions on Circuits and Systems for Video Technology}, 
  title={Overview of the H.264/AVC video coding standard}, 
  year={2003},
  volume={13},
  number={7},
  pages={560-576},
  doi={10.1109/TCSVT.2003.815165}}

@inproceedings{H264_2,
author = {Gary J. Sullivan and Pankaj N. Topiwala and Ajay Luthra},
title = {{The H.264/AVC Advanced Video Coding standard: overview and introduction to the fidelity range extensions}},
volume = {5558},
booktitle = {Applications of Digital Image Processing XXVII},
editor = {Andrew G. Tescher},
organization = {International Society for Optics and Photonics},
publisher = {SPIE},
pages = {454 -- 474},
year = {2004},
doi = {10.1117/12.564457}, 
}

@ARTICLE{H265,
  author={Sullivan, Gary J. and Ohm, Jens-Rainer and Han, Woo-Jin and Wiegand, Thomas},
  journal={IEEE Transactions on Circuits and Systems for Video Technology}, 
  title={Overview of the High Efficiency Video Coding (HEVC) Standard}, 
  year={2012},
  volume={22},
  number={12},
  pages={1649-1668},
  doi={10.1109/TCSVT.2012.2221191}}

@inproceedings{choi2020task,
  title={Task-aware quantization network for jpeg image compression},
  author={Choi, Jinyoung and Han, Bohyung},
  booktitle={Computer Vision--ECCV 2020: 16th European Conference, Glasgow, UK, August 23--28, 2020, Proceedings, Part XX 16},
  pages={309--324},
  year={2020},
  organization={Springer}
}

@inproceedings{ye2023accelir,
  title={AccelIR: Task-aware image compression for accelerating neural restoration},
  author={Ye, Juncheol and Yeo, Hyunho and Park, Jinwoo and Han, Dongsu},
  booktitle={Proceedings of the IEEE/CVF Conference on Computer Vision and Pattern Recognition},
  pages={18216--18226},
  year={2023}
}

@inproceedings{xie2019source,
  title={Source compression with bounded dnn perception loss for iot edge computer vision},
  author={Xie, Xiufeng and Kim, Kyu-Han},
  booktitle={The 25th Annual International Conference on Mobile Computing and Networking},
  pages={1--16},
  year={2019}
}

@inproceedings{xie2022bandwidth,
  title={Bandwidth-aware adaptive codec for dnn inference offloading in iot},
  author={Xie, Xiufeng and Zhou, Ning and Zhu, Wentao and Liu, Ji},
  booktitle={European Conference on Computer Vision},
  pages={88--104},
  year={2022},
  organization={Springer}
}

@ARTICLE{10585292,
  author={Salamah, Ahmed H. and Zheng, Kaixiang and Ye, Linfeng and Yang, En-Hui},
  journal={IEEE Journal on Selected Areas in Information Theory}, 
  title={JPEG Compliant Compression for DNN Vision}, 
  year={2024},
  volume={5},
  number={},
  pages={520-533},
  doi={10.1109/JSAIT.2024.3422011}}

@article{liu2025efficient,
  title={An Efficient Adaptive Compression Method for Human Perception and Machine Vision Tasks},
  author={Liu, Lei and Chen, Zhenghao and Hu, Zhihao and Xu, Dong},
  journal={arXiv preprint arXiv:2501.04329},
  year={2025}
}

@article{JEONG20231,
title = {An overhead-free region-based JPEG framework for task-driven image compression},
journal = {Pattern Recognition Letters},
volume = {165},
pages = {1-8},
year = {2023},
issn = {0167-8655},
doi = {https://doi.org/10.1016/j.patrec.2022.11.020},
author = {Seonghye Jeong and Seongmoon Jeong and Simon S. Woo and Jong Hwan Ko}
}

@ARTICLE{9462938,
  author={Mukherjee, Mandovi and Mudassar, Burhan Ahmad and Lee, Minah and Lee, Edward and Mukhopadhyay, Saibal},
  journal={IEEE Sensors Journal}, 
  title={Energy Efficient Pixel-Parallel Read-Out Circuits for Digital Image Sensors Using Cross-Layer Pixel Depth Control}, 
  year={2022},
  volume={22},
  number={12},
  pages={11317-11327},
  doi={10.1109/JSEN.2021.3091877}}

@inproceedings{ma2023leca,
  title={Leca: In-sensor learned compressive acquisition for efficient machine vision on the edge},
  author={Ma, Tianrui and Boloor, Adith Jagadish and Yang, Xiangxing and Cao, Weidong and Williams, Patrick and Sun, Nan and Chakrabarti, Ayan and Zhang, Xuan},
  booktitle={Proceedings of the 50th Annual International Symposium on Computer Architecture},
  pages={1--14},
  year={2023}
}

@inproceedings{finateu20205,
  title={5.10 a 1280$\times$ 720 back-illuminated stacked temporal contrast event-based vision sensor with 4.86 $\mu$m pixels, 1.066 GEPS readout, programmable event-rate controller and compressive data-formatting pipeline},
  author={Finateu, Thomas and Niwa, Atsumi and Matolin, Daniel and Tsuchimoto, Koya and Mascheroni, Andrea and Reynaud, Etienne and Mostafalu, Pooria and Brady, Frederick and Chotard, Ludovic and LeGoff, Florian and others},
  booktitle={2020 IEEE International Solid-State Circuits Conference-(ISSCC)},
  pages={112--114},
  year={2020},
  organization={IEEE}
}

@article{ragusa2024combining,
  title={Combining compressed sensing and neural architecture search for sensor-near vibration diagnostics},
  author={Ragusa, Edoardo and Zonzini, Federica and Gastaldo, Paolo and De Marchi, Luca},
  journal={IEEE Transactions on Industrial Informatics},
  year={2024},
  publisher={IEEE}
}

@inproceedings{kumar202365,
  title={A 65 nm 2.02 mw 50 mbps direct analog to MJPEG converter for video sensor nodes using low-noise switched capacitor MAC-Quantizer with automatic calibration and sparsity-aware ADC},
  author={Kumar, K Gaurav and Barik, Gourab and Chatterjee, Baibhab and Bose, Sumon and Maity, Shovan and Sen, Shreyas},
  booktitle={2023 IEEE Custom Integrated Circuits Conference (CICC)},
  pages={1--2},
  year={2023},
  organization={IEEE}
}

@article{kawahito1997cmos,
  title={A CMOS image sensor with analog two-dimensional DCT-based compression circuits for one-chip cameras},
  author={Kawahito, Shoji and Yoshida, Makoto and Sasaki, Masaaki and Umehara, Keijiro and Miyazaki, Daisuke and Tadokoro, Yoshiaki and Murata, Kenji and Doushou, Shirou and Matsuzawa, Akira},
  journal={IEEE Journal of solid-state circuits},
  volume={32},
  number={12},
  pages={2030--2041},
  year={1997},
  publisher={IEEE}
}

@inproceedings{AJPEG,
  author={Wan, Rentao and Xu, Yichen and Jee, Dong-Woo and Seok, Mingoo},
  booktitle={2025 IEEE Custom Integrated Circuits Conference (CICC)}, 
  title={AJPEG: A 26.4-pJ/pixel, 252-fps, 128x128 Image Sensor with an In-Sensor Analog DCT Processor for Data Compression}, 
  year={2025},
  volume={},
  number={},
  pages={1-2} }

@article{reddy2021pragmatic,
  title={Pragmatic image compression for human-in-the-loop decision-making},
  author={Reddy, Sid and Dragan, Anca and Levine, Sergey},
  journal={Advances in Neural Information Processing Systems},
  volume={34},
  pages={26499--26510},
  year={2021}
}

@article{shao2018crowdhuman,
  title={Crowdhuman: A benchmark for detecting human in a crowd},
  author={Shao, Shuai and Zhao, Zijian and Li, Boxun and Xiao, Tete and Yu, Gang and Zhang, Xiangyu and Sun, Jian},
  journal={arXiv preprint arXiv:1805.00123},
  year={2018}
}

@inproceedings{zhang2017citypersons,
  title={Citypersons: A diverse dataset for pedestrian detection},
  author={Zhang, Shanshan and Benenson, Rodrigo and Schiele, Bernt},
  booktitle={Proceedings of the IEEE conference on computer vision and pattern recognition},
  pages={3213--3221},
  year={2017}
}

@inproceedings{ess2008mobile,
  title={A mobile vision system for robust multi-person tracking},
  author={Ess, Andreas and Leibe, Bastian and Schindler, Konrad and Van Gool, Luc},
  booktitle={2008 IEEE conference on computer vision and pattern recognition},
  pages={1--8},
  year={2008},
  organization={IEEE}
}

@article{arai1988fast,
  title={A fast DCT-SQ scheme for images},
  author={Arai, Yukihiro and Agui, Takeshi and Nakajima, Masayuki},
  journal={IEICE TRANSACTIONS (1976-1990)},
  volume={71},
  number={11},
  pages={1095--1097},
  year={1988},
  publisher={The Institute of Electronics, Information and Communication Engineers}
}

@ARTICLE{feig1992fast,
  author={Feig, E. and Winograd, S.},
  journal={IEEE Transactions on Signal Processing}, 
  title={Fast algorithms for the discrete cosine transform}, 
  year={1992},
  volume={40},
  number={9},
  pages={2174-2193},
  doi={10.1109/78.157218}}

@INPROCEEDINGS{Barua2019Navigation,
  author={Barua, Bhaskar and Gomes, Clarence and Baghe, Shubham and Sisodia, Jignesh},
  booktitle={2019 International Conference on Intelligent Computing and Control Systems (ICCS)}, 
  title={A Self-Driving Car Implementation using Computer Vision for Detection and Navigation}, 
  year={2019},
  volume={},
  number={},
  pages={271-274},
  doi={10.1109/ICCS45141.2019.9065627}}

@inproceedings{tabelini2021keep,
  title={Keep your eyes on the lane: Real-time attention-guided lane detection},
  author={Tabelini, Lucas and Berriel, Rodrigo and Paixao, Thiago M and Badue, Claudine and De Souza, Alberto F and Oliveira-Santos, Thiago},
  booktitle={Proceedings of the IEEE/CVF conference on computer vision and pattern recognition},
  pages={294--302},
  year={2021}
}

@inproceedings{wang2023yolov7,
    title={YOLOv7: Trainable bag-of-freebies sets new state-of-the-art for real-time object detectors},
    author={Wang, Chien-Yao and Bochkovskiy, Alexey and Liao, Hong-Yuan Mark},
    booktitle={Proceedings of the IEEE/CVF conference on computer vision and pattern recognition},
    pages={7464--7475},
    year={2023}
}

@inproceedings{mullapudi2018hydranets,
  title={Hydranets: Specialized dynamic architectures for efficient inference},
  author={Mullapudi, Ravi Teja and Mark, William R and Shazeer, Noam and Fatahalian, Kayvon},
  booktitle={Proceedings of the IEEE conference on computer vision and pattern recognition},
  pages={8080--8089},
  year={2018}
}

@article{bojarski2016end,
  title={End to end learning for self-driving cars},
  author={Bojarski, Mariusz and Del Testa, Davide and Dworakowski, Daniel and Firner, Bernhard and Flepp, Beat and Goyal, Prasoon and Jackel, Lawrence D and Monfort, Mathew and Muller, Urs and Zhang, Jiakai and others},
  journal={arXiv preprint arXiv:1604.07316},
  year={2016}
}

@inproceedings{liu2018deepn,
  title={DeepN-JPEG: A deep neural network favorable JPEG-based image compression framework},
  author={Liu, Zihao and Liu, Tao and Wen, Wujie and Jiang, Lei and Xu, Jie and Wang, Yanzhi and Quan, Gang},
  booktitle={Proceedings of the 55th annual design automation conference},
  pages={1--6},
  year={2018}
}

@inproceedings{shimauchi2004jpeg_resolution,
  title={JPEG based image compression with adaptive resolution conversion system},
  author={Shimauchi, Kazuhiro and Ogawa, Masahiro and Taguchi, Akira},
  booktitle={2004 IEEE International Symposium on Circuits and Systems (ISCAS)},
  volume={5},
  pages={V--V},
  year={2004},
  organization={IEEE}
}

@inproceedings{bannon2019computer,
  title={Computer and redundancy solution for the full self-driving computer},
  author={Bannon, Pete and Venkataramanan, Ganesh and Sarma, Debjit Das and Talpes, Emil},
  booktitle={2019 IEEE Hot Chips 31 Symposium (HCS)},
  pages={1--22},
  year={2019},
  organization={IEEE Computer Society}
}

@article{coffin2019building,
  title={Building vehicle autonomy: Sensors, semiconductors, software and US competitiveness},
  author={Coffin, David and Oliver, Sarah and VerWey, John},
  journal={Office of Industries Working Paper ID-063},
  year={November 2019}
}

@article{wu2022yolop,
  title={Yolop: You only look once for panoptic driving perception},
  author={Wu, Dong and Liao, Man-Wen and Zhang, Wei-Tian and Wang, Xing-Gang and Bai, Xiang and Cheng, Wen-Qing and Liu, Wen-Yu},
  journal={Machine Intelligence Research},
  volume={19},
  number={6},
  pages={550--562},
  year={2022},
  publisher={Springer}
}

@article{yu2018bdd100k,
  title={Bdd100k: A diverse driving video database with scalable annotation tooling},
  author={Yu, Fisher and Xian, Wenqi and Chen, Yingying and Liu, Fangchen and Liao, Mike and Madhavan, Vashisht and Darrell, Trevor and others},
  journal={arXiv preprint arXiv:1805.04687},
  volume={2},
  number={5},
  pages={6},
  year={2018}
}

@incollection{boncelet2009image,
  title={Image noise models},
  author={Boncelet, Charles},
  booktitle={The essential guide to image processing},
  pages={143--167},
  year={2009},
  publisher={Elsevier}
}

@article{boyat2015review,
  title={A review paper: noise models in digital image processing},
  author={Boyat, Ajay Kumar and Joshi, Brijendra Kumar},
  journal={arXiv preprint arXiv:1505.03489},
  year={2015}
}

@article{damaj2022future,
  title={Future trends in connected and autonomous vehicles: Enabling communications and processing technologies},
  author={Damaj, Issam W and Yousafzai, Jibran K and Mouftah, Hussein T},
  journal={IEEE Access},
  volume={10},
  pages={42334--42345},
  year={2022},
  publisher={IEEE}
}

@book{yang2020nature,
  title={Nature-inspired optimization algorithms},
  author={Yang, Xin-She},
  year={2020},
  publisher={Academic Press}
}

@misc{DS90UB953,
  title={DS90UB953-Q1 FPD-Link III 4.16Gbps Serializer With CSI-2 Interface for 2.3MP/60fps
Cameras, RADAR, and Other Sensors},
  author={Texas Instruments},
  year={2017},
  url = {https://www.ti.com/lit/ds/symlink/ds90ub953-q1.pdf}
}

@misc{DS90UB954,
  title={DS90UB954-Q1 Dual 4.16 Gbps FPD-Link III Deserializer Hub With MIPI CSI-2 Outputs
for 2MP/60fps Cameras and RADAR},
  author={Jagdish, Kaushik and Vikas, Choudhary},
  year={2023},
  url = {https://www.ti.com/lit/ds/symlink/ds90ub954-q1.pdf}
}
\end{document}